\documentstyle[prb,aps,preprint,epsfig]{revtex}
\tightenlines 
\newcommand{\be}{\begin{equation}}
\newcommand{\ee}{\end{equation}}

\begin{document}

\title{A detector of small harmonic displacements based on two coupled microwave cavities}

\author{Ph. Bernard} 
\address{CERN, CH--1211, Geneva 23, Switzerland}
\author{G. Gemme\thanks{Corresponding author. E--mail: gianluca.gemme@ge.infn.it} \and R. Parodi}
\address{INFN--Sezione di Genova, via Dodecaneso 33, I--16146 Genova, Italy}
\author{E. Picasso}
\address{Scuola Normale Superiore, Piazza dei Cavalieri 7, I--56126, Pisa, Italy}

\maketitle

\begin{abstract}
The design and test of a detector of small harmonic displacements is presented. The detector is based on the principle of the parametric conversion of power between the resonant modes of two superconducting coupled microwave cavities. The work is based on the original ideas of Bernard, Pegoraro, Picasso and Radicati, who, in 1978, suggested that superconducting coupled cavities could be used as sensitive detectors of gravitational waves, and on the work of Reece, Reiner and Melissinos, who, {  in 1984}, built a detector of this kind. They showed that an harmonic modulation of the cavity length $\ell$ produced an energy transfer between two modes of the cavity, provided that the frequency of the modulation was equal to the frequency difference of the two modes. They achieved a sensitivity to fractional deformations of $\delta \ell / \ell \approx 10^{-17}$ Hz$^{-1/2}$.
We repeated the Reece, Reiner and Melissinos experiment, and with an improved experimental configuration and better cavity quality, increased the sensitivity to $\delta \ell / \ell \approx 10^{-20}$ Hz$^{-1/2}$. 
In this paper the basic principles of the device are discussed and the experimental technique is explained in detail. Possible future developments, aiming at gravitational waves detection, are also outlined.
\end{abstract}

\section{Introduction}
\label{sec:intro}

The detection of weak forces acting on macroscopic bodies often entails the 
measurement of extremely small displacements of these bodies or their 
boundaries. This is the case in the attempts to detect gravitational waves 
(g.w.) or in the search for possible new long range interactions. In 1978 
Bernard, Pegoraro, Picasso and Radicati (BPPR) suggested that superconducting 
coupled cavities could be used as a sensitive detector of gravitational 
effects, through the coupling of the gravitational wave with the cavity walls and with the 
stored electromagnetic field. It 
has been shown that the principle that underlies this detector is analogous 
to the one used in parametric processes and in particular in frequency 
converters, i.e. in a device which converts energy from a reference 
frequency to a signal at a different frequency as a consequence of the time 
variation of a parameter of the system\cite{bppr1,bppr2}. In more detail the rf 
superconducting coupled cavities detector consists of an electromagnetic 
resonator, with two levels whose frequencies $\omega_{s}$ and 
$\omega_{a}$ are both much larger than the angular frequency $\Omega $ 
of the g.w. and satisfy the resonance condition $| \omega_{a} - \omega_{s} | = \Omega $. 
In the scheme suggested by Bernard et al. the two levels are obtained by coupling two 
identical high frequency resonators. The angular frequency $\omega_{s}$ 
is the frequency of the level symmetrical in the fields of the two 
resonators, and $\omega_{a}$ is that of the antisymmetrical one; if initially some energy $U$ is stored
in, say, the symmetrical level, leaving empty the antisymmetrical one, 
the g.w. can induce a transition between the two levels, 
provided that a suitable geometric configuration of the detector is chosen. In fact the interaction between the g.w. and the detector is characterized by a transfer of {\em energy} and of {\em angular momentum}. Since the elicity of the g.w. (i.e. the angular momentum along its direction of propagation) is 2, it can induce a transition between the two levels provided their angular momenta, along the direction of the g.w., differ by 2. This can be achieved by putting the two resonators at right angle or by a suitable polarization of the e.m. field inside the cavity.
 
The power $P$ exchanged between the g.w. and the detector is proportional to the incoming 
gravitational flux $\Phi_{gw}$, to the electromagnetic energy $U$ stored in the cavity, and it 
depends upon the mechanical properties of the detector and the spatial symmetries of the stored electromagnetic field; 
it can be written as:
\be
\label{eq:fond}
P = \frac{1}{4} h^2 |\Lambda(\Omega)|^2 |{\mathcal C}|^2 \Omega U Q_L
\ee
where $h$ is the adimensional amplitude of the g.w., usually defined in terms of the incoming gravitational flux as\cite{thorne}
\be
h = \frac{1}{\Omega}\left ( 32 \pi \frac{G}{c^3} \Phi_{gw} \right )^{1/2}
\ee
$\Omega$ is the g.w. angular frequency and $Q_L$ is the loaded electromagnetic quality factor of the microwave cavity. {  $\Lambda(\Omega)$ is the mechanical transfer function of the detector}. If we apply a simple model in which the detector mechanical properties are approximated by those of a simple harmonic oscillator with proper angular frequency $\omega_m$ and quality factor $Q_m$ we obtain\cite{bppr2}
\be
\label{eq:lambda}
|\Lambda(\Omega)|^2 = \frac{\Omega^4}{\left(\Omega^2-\omega_m^2\right)^2+\frac{\Omega^2\omega_m^2}{Q_m^2}}
\ee 
We see that for $\Omega \gg \omega_m$, $|\Lambda(\Omega)|^2 \approx 1$, while near a mecanical resonance $|\Lambda(\Omega)|^2 \approx Q_m^2$; in the following, since the working frequency of our detector is chosen to be $\Omega / 2 \pi \approx 1$ MHz, we shall take in our formulas $|\Lambda(\Omega)| \approx 1$.

${\mathcal C}$ is a complex coefficient, depending on the symmetries of the fields of the electromagnetic modes coupled by the g.w. and is defined so that $|{\mathcal C}| \leq 1$. Obviously a clever choice of field symmetries has to be done in order to have 
$|{\mathcal C}| \approx 1$ when designing a detector. The discussion on field symmetries and their influence on g.w. detection is done elsewhere\cite{bgpp3}. In the following we shall suppose that the system is designed to have $|{\mathcal C}| = 1$.

In 1984 Reece, Reiner and Melissinos (RRM), built a detector in the 10 GHz frequency range, 
using two identical cylindrical cavities mounted {\em end--to--end} and coupled via a small 
circular aperture in their common endwall, and tested it as a transducer of 
harmonic mechanical motion\cite{rrm1,rrm2}. 
In order to measure the sensitivity limit of the detector, one of 
its walls was excited by an harmonic perturbation (by a piezoelectric); with 
a quality factor of the superconducting cavity  $Q = 3 \times 10^{8}$ 
and a stored electromagnetic energy $U \approx 0.1$ mJ they 
showed a sensitivity to relative deformations $\delta \ell / \ell \approx 10^{-17}$ Hz$^{-1/2}$,
with $\Omega / 2 \pi \approx 1$ MHz.
We repeated the RRM experiment and, with 
a quality factor $Q \approx 10^{8}$  and a stored energy $U = 1.8$ J, improved its sensitivity by a factor 
$10^{3}$, thus reaching a sensitivity to harmonic displacements of 
the order $\delta \ell / \ell \approx 10^{-20}$ (Hz)$^{-1/2}$ at $\Omega / 2 \pi \approx 1\,$ MHz. With these figures this 
detector could be an interesting candidate for the detection of gravitational waves or in the search of long range interactions of weak intensity.

\section{Basic principles of parametric power conversion}
\label{sec:basic}

A parametric converter is a nonlinear device which transfers energy from a 
reference frequency to a signal with different frequency, utilizing a 
nonlinear parameter of the system, or a parameter that can be 
varied as a function of time by applying a suitable signal. The time varying 
parameter may be electrical or mechanical; in the latter case the device 
acts as a transducer of mechanical displacements.

The basic equations describing the parametric converter are the Manley--Rowe 
relations\cite{manrov,louisell}. They are a set of power conservation 
relations that are extremely useful in evaluating the performance of a 
parametric device. We will not derive here the complete Manley--Rowe 
relations, but we will just describe the fundamental ideas upon which the 
parametric converter behavior is based.

Let us consider a physical system which can exist in two distinct energy 
levels. To fix our ideas let us take as an example a system of two identical 
coupled resonant cavities like those proposed by BPPR. If the resonant 
angular frequency of the unperturbed single cell is $\omega _0$, then the 
frequency spectrum of the coupled system consists of two levels at angular
frequencies $\omega _s$ and $\omega _a$, where $\omega_s = \omega_0 -\delta \omega$ 
and $\omega_a = \omega_0 + \delta \omega $, and 
$2 \,\delta \omega / \omega_0 = K \ll 1$, is the coupling 
coefficient of the system.

We can store some electromagnetic energy in the system, say at angular frequency $\omega_s$; 
if some external harmonic perturbation at angular frequency $\Omega = \omega_a - \omega_s$ induces 
the time variation of one system parameter we can have the transfer of energy 
between the two energy levels. The external perturbation can be an harmonic 
modulation of the cavity end wall by a piezoelectric crystal, which causes the variation of the system 
reactance, or a gravitational wave deforming the cavity walls.

Let us call $P$ the total power absorbed by the device. We can write
\be
P = P_s + P_a = \dot N_s \hbar \omega _s + \dot N_a \hbar \omega _a
\ee
where $\dot N_i $ is the time variation of the number of photons in 
level {\em i}. If we assume that the total number of photons in the 
systems $N = N_s + N_a$ is conserved we have
\be
\dot N_a = - \dot N_s = \dot N
\ee

The total power absorbed by the system is therefore
\be
P = \dot N \hbar \left( \omega _a - \omega _s  \right)
\ee
and the power transferred to the upper level is given by
\be
\label{eq:pex}
P_a = \dot N \hbar \omega_a = \frac{P}{\hbar(\omega_a - \omega_s)}\hbar \omega_a =
\frac{\omega_a}{\Omega} P
\ee
The amplification factor $\omega _a/ \Omega $ is characteristic of parametric frequency converters.

To get deeper insight into eq. \ref{eq:pex} we have to know the exact expression of 
$P$; in fact the power exchanged between the external perturbation 
and the system is proportional to the square of the fractional change of the 
time varying system parameter. If we denote this quantity with $h$, 
and use eq. \ref{eq:fond} for the exchanged power (with $|\Lambda| = |{\mathcal C}| =1$) we can easily derive the expression for the power 
transferred to the initially empty level
\be
\label{eq:pout}
P_a = \frac{1}{4} h^2 \omega _a U_s Q_{sL} 
\ee
In the last equation $U_s$ is the energy stored in level 
$s$, $Q_{sL}$ is the loaded quality factor of the cavity excited in the symmetric mode. 

The quantity $h$ above may represent the fractional change of the 
system reactance, or of the system length being harmonically modulated 
($\delta \ell / \ell$), or the dimensionless amplitude of the g.w.

Since $P_a$ is proportional to the electromagnetic quality 
factor, superconducting resonant cavities should be employed to achieve 
maximum sensitivity.

\section{Sensitivity limits}
\label{sec:sens}

Equation \ref{eq:pout} could be a good starting point for a detailed discussion of the 
theoretical sensitivity limits of the detector. Let us choose a configuration 
very similar to the RRM experiment with two cylindrical niobium cavities 
coupled through a small circular aperture on the axis and let the operating mode be 
the TE$_{011}$ at 3 GHz, which, due to the coupling, splits into a 
symmetrical and an antisymmetrical mode respectively at angular frequency $\omega_{s}$
and $\omega_{a}$. The system is designed so that mode separation is about 1 MHz 
(see section \ref{sec:design} for more details on system design). 
For our geometry, we can calculate the relation between the maximum energy 
that can be stored in a superconducting niobium cavity (due to the thermodinamical
critical field $H_c$) and the frequency of the electromagnetic field\cite{nota1}.
The temperature dependence of the critical magnetic field is given by
\be
H_c(T) = H_c(0)\left[1-\left(\frac{T}{T_c}\right)^2\right]
\ee
For niobium $T_c = 9.2$ K and $H_c(T=0) = 0.2$ T. 

Let us suppose to work at $T = 1.8$ K, with a stored energy corresponding to a maximum
surface magnetic field $H_{max} = 0.75 \times H_c(T=1.8)$. 
We have $H_c(T=1.8) \approx 0.19$ T, and $H_{max} \approx 0.14$ T. 
We find for a pill--box cavity at 3 GHz
\be
\label{eq:umax}
\omega U_{max} \approx 10^{11} \mathrm{Joule\,\frac{rad}{sec}}
\ee

If we define the minimum measurable $h$ per unit bandwidth, $h_{min}$, as the perturbation value which induces a parametric conversion of power equal to the noise power spectral density in the system, ${\mathcal P}_{min}$, we find, deriving $h$ from eq.  \ref{eq:pout} and using eq. \ref{eq:umax}:
\be
\label{eq:hmin}
h_{min} \approx \left ( \frac{{\mathcal P}_{min}}{\omega U_{max} Q_L}\right )^{1/2}
\ee
{  Bearing in mind that in a cavity the stored energy is proportional to the cavity volume (keeping constants the fields, i.e. the incident power and the coupling coefficients) and that the resonant angular frequency is inversely proportional to the cavity linear dimension}, we find that $U_{max} \propto 1 / \omega^3$, so that $\omega^3 U_{max} =$ const., eq. \ref{eq:hmin} can be put in the form:
\be
h_{min} \approx 3 \times 10^{-22} \left ( \frac{\omega / 2\pi}{1 \mathrm{GHz}}\right ) 
\left ( \frac{{\mathcal P}_{min}}{10^{-22}\,\mathrm{Watt/Hz}}\right )^{1/2} 
\left ( \frac{10^{10}}{Q_L} \right )^{1/2}\mathrm{Hz}^{-1/2}
\ee

From eq. \ref{eq:hmin} is apparent that to get better sensitivity lower operating
frequencies and higher quality factors are preferred. 

\section{Electromagnetic noise at angular frequency $\omega_a$}
\label{sec:em_noise}

To operate our device we have to feed microwave power into one resonant mode and then to perturb one system parameter at a frequency equal to the mode separation, in order to detect the energy transfer between the symmetric and the antisymmetric mode. Here we suppose that the initially full mode is the symmetrical one, and that this mode is the lower frequency one. This is very likely to be the case, but is not at all crucial for the following discussion.

If $P_i$ is the power input at angular frequency $\omega_s$, and the system bandwidth is determined by the quality factor $Q_L$ of the cavity, we have that the frequency distribution of the stored energy of the symmetric mode is given by the usual Breit--Wigner function, describing an oscillator's response to an harmonic driving force:
\be
U_{s}\left( \omega \right)  = \frac{\beta_1}{1+\beta_1}\frac{\frac{4P_i Q_L}{\omega _s }}{1 + Q_{L}^{2} \left( \frac{\omega}{\omega _s } - \frac{\omega _s }{\omega } \right)^2 }
\ee
From the above equation we can evaluate the symmetric mode power that can be extracted at the antisymmetric frequency, i.e. the tail of mode {\it s} at the frequency of mode {\it a}:
\be
{P}_{symm} \left( \omega _a \right) = \beta_2 \frac{\omega_a U_{s}\left( \omega_a \right)}{Q_0}
\approx \frac{\beta_1 \beta_2}{(1+\beta_1)^2}\frac{P_i}{Q_L^2} \left( \frac{\omega _a }{\Omega } \right)^2
\ee
where $\beta_1$ is the coupling coefficient of the {\em input} cavity port for the symmetric mode and $\beta_2$ is the coupling coefficient of the {\em output} cavity port for the antisymmetric mode; for optimum system sensitivity we should have $\beta_1 = \beta_2 = 1$.

\begin{figure}[p]
\begin{center} \mbox{\epsfig{file=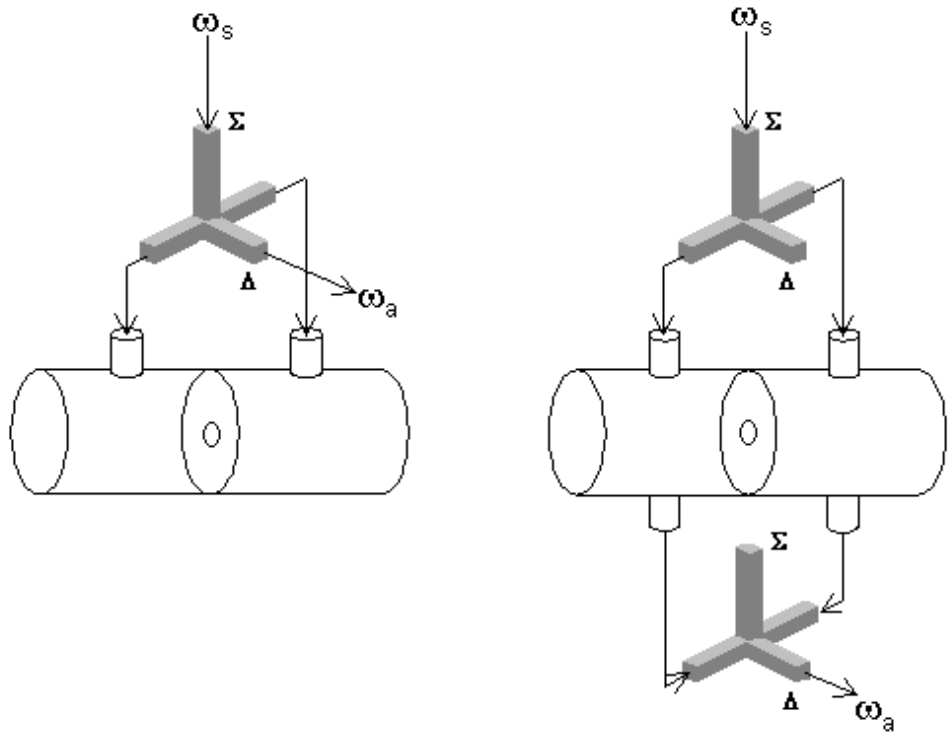}} \end{center}
\caption{Schematic view of the detection scheme in reflection (left) and in transmission (right)} 
\label{magictees} 
\end{figure}

It can easily be verified that this noise source at detection frequencies $\Omega \leq 10^5$ Hz lowers the detector sensitivity at an unacceptable level, even for high $Q$ superconducting cavities. A properly designed detection system can strongly reduce this problem: in fact what we want to do is to detect a small signal at angular frequency $\omega_a$, which is quite near to the frequency of the very large signal present at angular frequency $\omega_s$. Nevertheless we can make use of the fact that the two signals belong to resonant modes with very different {\em field symmetries}, the symmetric and antisymmetric mode. In fact we designed a detection system capable to discriminate the field parity, which attenuates the symmetric component at angular frequency $\omega_a$ by a factor $R$, while leaving unaffected the antisymmetric one. $R$ is a number which depends upon the details of the detection system; if the receiver does not discriminate the parity of the field at angular frequency $\omega_a$, $R \approx 1$. Discrimination of the parity would considerably reduce this value. 
RRM used a magic--tee to excite the symmetric mode through the $\Sigma$ port, and to detect the antisymmetric mode through the $\Delta$ port. With careful adjustments they obtained $R \approx 10^{-7}$. The detection system designed for our experiment, based on two magic--tees, one at the cavity input and one at the cavity output allowed to get $R \approx 10^{-14}$. For a schematic view see figure \ref{magictees}; experimental details are described in section \ref{sec:detec}.

\section{Mechanical and Electromagnetic Design of the Detector}
\label{sec:design}

The detector is built using two coupled rf cavities. The geometry of 
the resonators gives us the electromagnetic field needed for the detection 
of the wanted physical quantities.

Because the ultimate sensitivity of the detector at a given frequency is 
related to the quality factor and the stored energy of the resonator, a 
resonator geometry with high geometric magnetic factor $\Gamma $ is 
preferred, where
\begin{equation}
\label{eq21}
\Gamma = \omega _{0} \mu _{0} \frac{\int_{V} {H^{2}\,dV}}{\int_{S} {H^{2}\,dS}}
\end{equation}

To avoid rf electronic vacuum discharges and Fowler--Nordheim like non 
resonant electron loading, rf modes with vanishing electric field at the 
surface (TE modes) are preferred.

Early analysis of the best resonator shape suggested to use two spherical 
cavities coupled through an iris\cite{rfsc97}; nevertheless 
the dimension of the spherical resonator and the will to avoid unessential (at this stage) difficulties  
suggested us to use as a first step a conservative approach 
using two cylindrical cavities coupled through an axial iris.

The choice of the frequency was imposed by the maximum 
dimension for the resonator that can be housed in our standard test cryostat 
in a comfortable way; in our case the inner diameter is 300 mm giving us 
enough room for a 3 GHz resonator. 
We started our design with a couple of cylindrical resonators 
having the lenght equal to the diameter.
For this configuration we can compute all the relevant quantities, as 
geometric factor and frequencies, starting from analytical expressions for 
the fields, or using the Bethe\cite{bethe} perturbation theory for the 
evaluation of the coupling coefficient and mode separation.

\begin{figure}[p]
\begin{center} \mbox{\epsfig{file=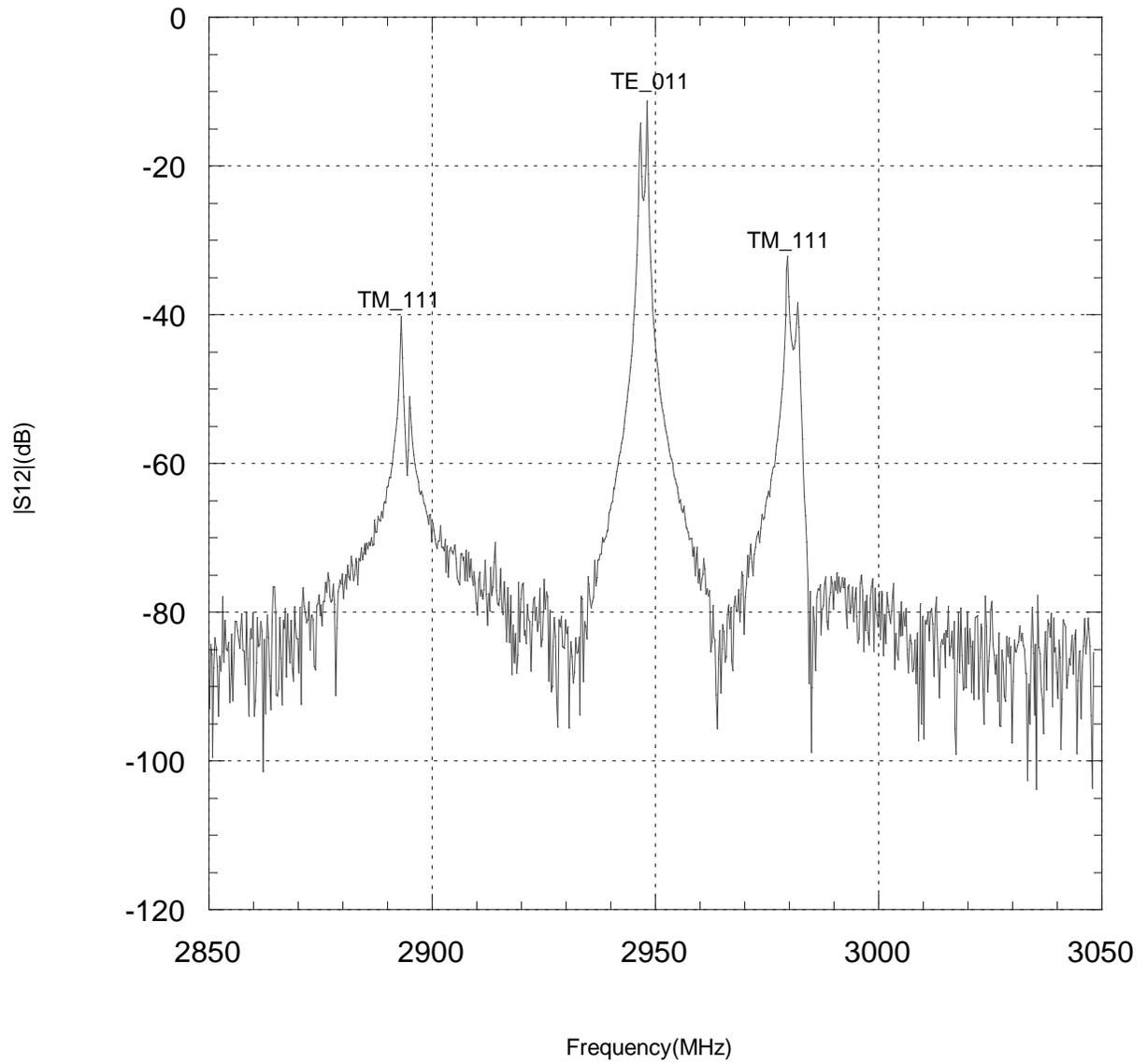}} \end{center}
\caption{TM--TE modes separation measured on the normal conducting cavity} 
\label{modesep} 
\end{figure}

Using a pill box like TE$_{011}$ geometry could be a little 
bit upsetting due to the degeneracy of the TE$_{01}$ and TM$_{11}$ modes; 
this problem is quite mild in our case: due to the very high $Q$ 
value of the superconducting cavities, any distortion of the cavity geometry 
will split the TE and TM modes avoiding unwanted interactions.
Nevertheless we would like to enhance the splitting to be sure to remove any 
possible interaction between the TE and TM resonances.
To do that we design the cavity with a little modified geometry substituting 
the straight end plates with a spherical segment; the effect of this 
modification is to move away by 50 MHz the TM$_{11}$ modes (see figure \ref{modesep}).

\begin{figure}[p]
\begin{center} \mbox{\epsfig{file=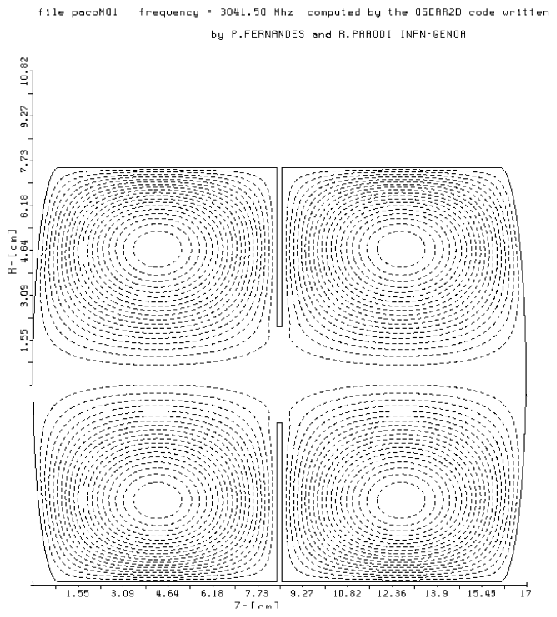,height=18cm}} \end{center}
\caption{2D layout of the coupled cavities with field lines of the symmetric mode} 
\label{couplres} 
\end{figure}

\begin{figure}[p]
\begin{center} \mbox{\epsfig{file=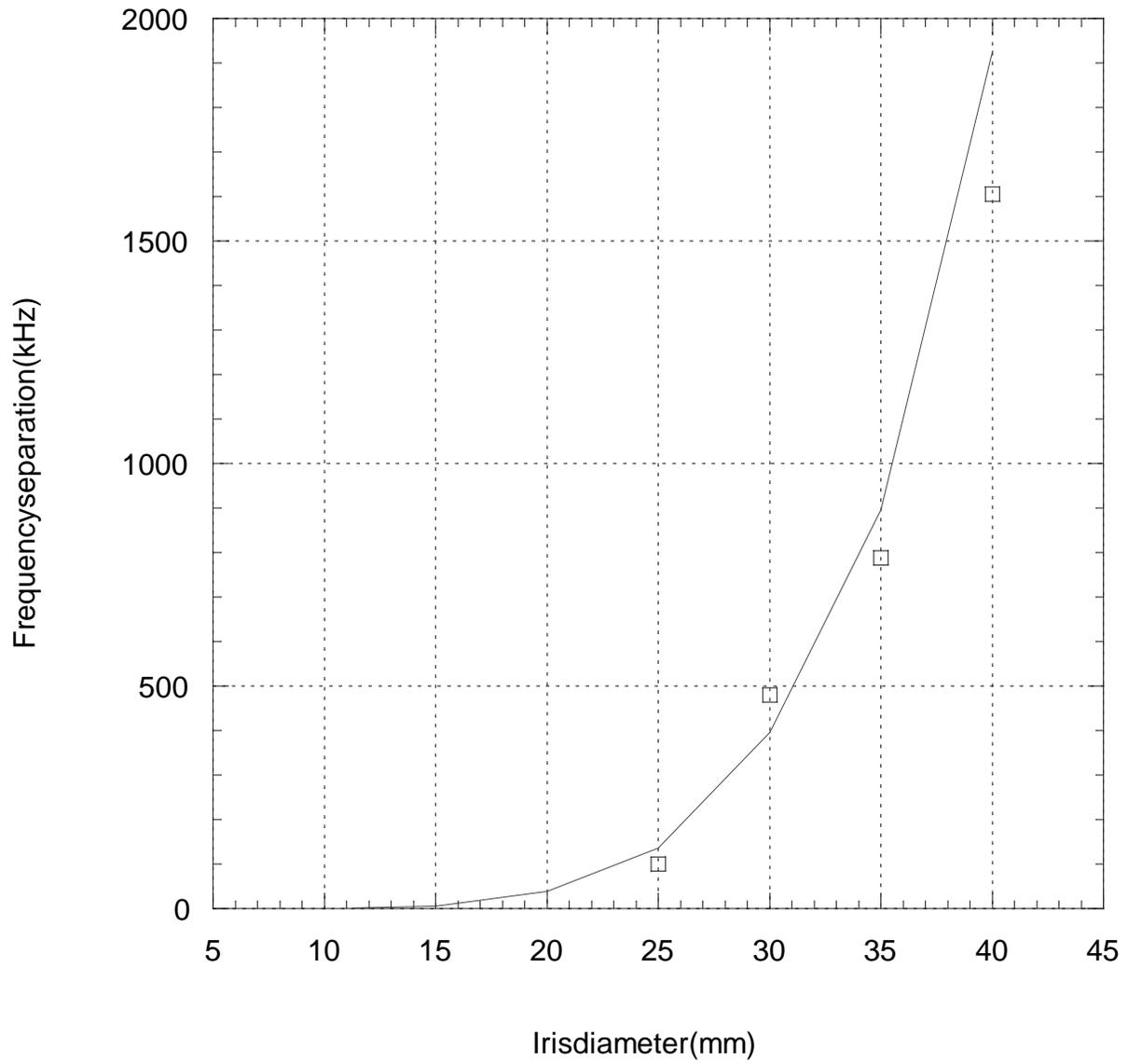}} \end{center}
\caption{Computed (solid line) and measured mode splitting vs. iris diameter. Measurements were made on a demountable copper model of the detector} 
\label{iris} 
\end{figure}

The selected geometry for the coupled resonators is reported on figure \ref{couplres}
with superimposed the magnetic field lines for the antisymmetric mode of 
the coupled resonators. With the coupling hole diameter used in the calculation (40 mm) the 
mode splitting is about 1.9 MHz. At the final design stage the hole diameter 
will be 35 mm to have $\Delta f \approx 1.3$ MHz.
The coupling iris is a circular hole on the cavity axis. The effect of the 
coupling gives a symmetric field distribution at the lower frequency and the 
antisymmetric at the higher frequency.
Running the resonator geometry with different openings of the coupling hole 
we got informations about the sensitivity of the coupling on the iris 
diameter (see figure \ref{iris}).

Finally the effect of the coupling ports on the resonant frequency is evaluated 
computing the field distribution and scattering matrix for the full 3D 
problem (including the coupling loops) in a frequency interval centered on 
the operating frequency of the detector. The full 3D simulations have 
been done with the HFSS package\cite{nota2}. 
The 3D simulation on the complete resonator confirmed the results of Oscar2D 
simulations\cite{oscar} both for the distribution of the fields of the resonant modes and for the 
coupling coefficient for a given iris diameter.

\begin{figure}[p]
\begin{center} \mbox{\epsfig{file=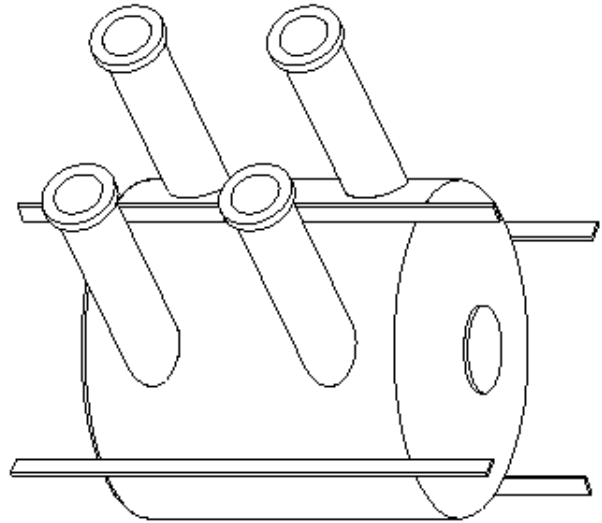}} \end{center}
\caption{Final design of the resonator} 
\label{paco} 
\end{figure}

On the basis of the results of the simulations the final design of the 
cavity was decided: the construction drawing of the resonator is shown 
in figure \ref{paco}.

Experimental results obtained with this device will be shown in section \ref{sec:results}.

\subsection{Mechanical Analysis}
\label{mech}

The mechanical behaviour of the resonator structure plays an important role 
in setting the ultimate sensitivity of the detector: as is shown in equation \ref{eq:fond}, the power absorbed by the device depends from its mechanical properties through the transfer function $\Lambda(\Omega)$ defined in equation \ref{eq:lambda}. Furthermore the thermal noise, due to the brownian motion of the cavity walls depends on the mechanical characteristics of the detector. If the cavity is approximated by a simple harmonic oscillator the wall fluctuates with a displacement spectral density that has the typical lorentzian shape\cite{ricci}:
\be
S_{xx} = \frac{4kT}{m \tau}\frac{1}{\left(\omega_m^2-\omega^2\right)^2+\left(\frac{\omega}{\tau}\right)^2}
\ee
where $\tau$ is the damping time of the oscillator. 
{  The sensitivity limit due to the thermal noise is roughly given by $h_{min} \approx \sqrt{S_{xx}}/L$, 
where $L$ is the detectors' length; setting $m \approx 10$ Kg, $L \approx 0.3$ m, 
$Q_m = \omega_m \tau \approx 10^3$, we find, at $\Omega = 1$ MHz and at $T = 1.8$ K, 
$h_{min} \approx 10^{-21}$ Hz$^{1/2}$, an order of magnitude better than the limit imposed by our detection electronics at the same frequency.} 

The knowledge of the deformation of the cavity walls (for an 
harmonic distortion of given amplitude and frequency) allows us to have a 
calibration of the detector when driven by a piezoelectric crystal in the test runs.

To estimate the modal spectrum of the resonator and the response to a 1 mm 
displacement of a portion of the end wall we used the ANSYS mechanical 
analysis package\cite{nota3}.

The frequencies of the first 19 mechanical modes of the cavity, computed 
using a full 3D model are reported on table I. The modes from 4 to 11 are 
all doubly degenerate.

\begin{figure}[p]
\begin{center} \mbox{\epsfig{file=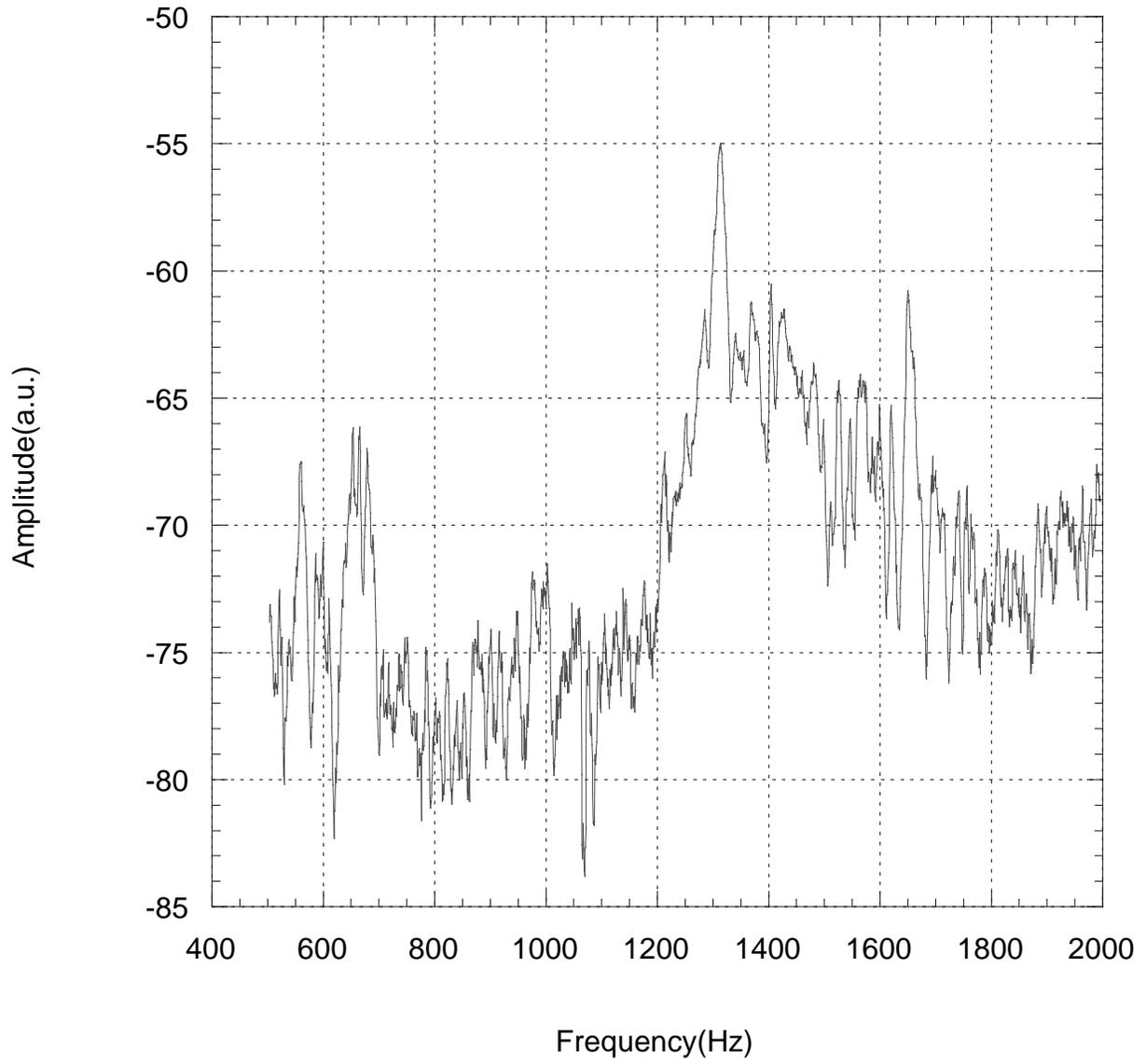}} \end{center}
\caption{Mechanical spectrum measurement (see section \ref{mech} for details)} 
\label{mismecc} 
\end{figure}

The mechanical spectrum of the detector has been mesured in the frequency range 500--2000 Hz
using a piezoelectric crystal fixed to a cavity end wall and driven by a sinusoidal signal of fixed amplitude
(1 V) and varying frequency. Since electromagnetic energy is fed into the cavity by a voltage controlled
oscillator locked to the cavity resonant frequency by a feedback signal proportional to the displacement 
of the cavity frequency, coming from the IF port of an harmonic mixer, we could use the feedback signal to detect the cavity
response to the mechanical excitation. A small portion ($\approx 1 \%$)
of the feedback signal was sent to a Stanford Research SR780 FFT Signal Analyzer, operating in swept sine mode and providing the excitation signal for the piezoelectric crystal as well.
The measurement results are shown in figure \ref{mismecc}, showing good agreement with the 
calculated mode distribution. The quality factor of the modes could be estimated from the data of figure \ref{mismecc} and is $Q_m \leq  1000$. Above 2000 Hz essentially no isolated mechanical mode was detected. 

After the modal analysis the study of the mechanical response to a fixed 
amplitude sinusoidal displacement of one end plate was performed using a 2D 
model to cope with the tiny mesh size needed at the higher frequencies to 
get meaningful results.

This simulations gave us informations about the transition frequency from the insulated mechanical resonances regime to the elastic 
continuum regime, giving again useful information about the influence of the 
thermal noise of the cavity walls on the detector ultimate sensitivity\cite{bgpp2}.

The harmonic analysis, performed at four frequencies 1 KHz, 10 KHz, 100 KHz 
and 1 MHz, showed a wall distortion decreasing with increasing frequency.

\section{rf Control and Detection System Design}
\label{sec:detec}

\begin{figure}[p]
\begin{center} \mbox{\epsfig{file=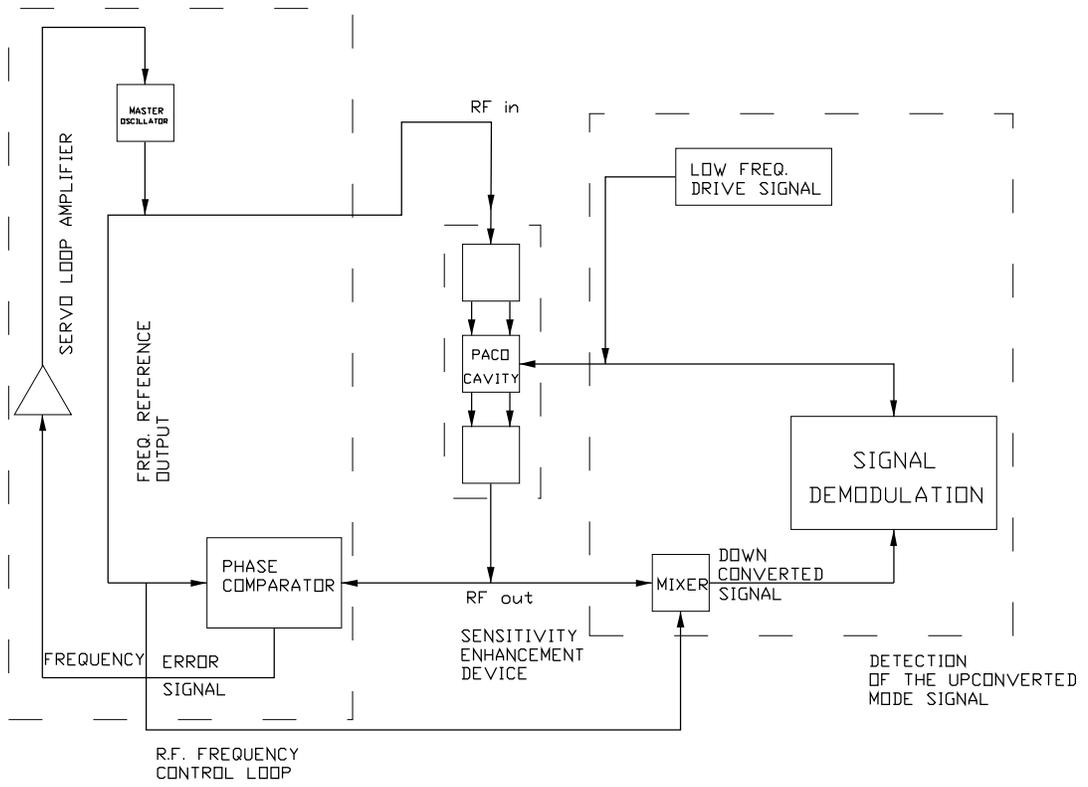}} \end{center}
\caption{rf system general layout} 
\label{general} 
\end{figure}

The three main functions of rf control and measurement system of the 
experiment are shown in figure \ref{general}:
\begin{enumerate}
\item{The first task of the system is to lock the rf frequency of the master 
oscillator to the resonant frequency of the symmetric mode of the cavity and to keep 
constant the energy stored in the mode. 
The frequency lock of the master oscillator to the cavity mode is necessary 
to fill in energy in the fundamental mode of the cavity. 
The reduction of fluctuations of the stored energy to less than 100 ppm 
greatly reduces the possibility of ponderomotive effects due to the 
radiation pressure of the electromagnetic field on the cavity walls and helps to minimize 
the contribution of the mechanical perturbations of the resonator to the 
output noise.
The frequency lock allows also to design a detection scheme insensitive to 
fluctuations of the resonant frequency of the two cavities forming our 
detector.}
\item{The second task is to increase the detector's sensitivity by driving the 
coupled resonators purely in the symmetric mode and receiving only the rf power up--converted to the antisymmetric mode by the perturbation of the cavity walls. This goal can be 
obtained by rejecting the signal at the symmetric mode frequency taking advantage of the 
symmetries in the field distribution of the two modes.
Our system, despite of some additional complexity, guarantees the following improvements 
over the one used in previous experiments:
\begin{itemize}
\item{a better rejection of the phase noise of the master 
oscillator obtained using the sharp resonance ($Q=10^{10}$) of the resonator 
as a filter;}
\item{a better insulation of the drive and detector ports obtained by 
using separate drive and detection arms of the rf system;}
\item{the possibility of an independent adjustment of the phase lag in 
the two arms giving a better magic--tee insulation at the operating 
frequencies;}
\item{a greater reliability for the frequency amplitude loop using the 
transmitted power, instead of the reflected, coming from the cavity.}
\end{itemize}
}
\item{The third task is the detection of the up--converted signal achieving the 
detector sensitivity limit set by the contribution of the noise sources\cite{bgpp2}. 
Slow pressure fluctuations on the cooling bath, hydrostatic pressure 
variations due to the changes in the helium level, pressure radiation, and so on, will 
change in the same way the resonant frequency of both modes.
Using a fraction of the main oscillator power as local oscillator for the 
detection mixer, the detection system becomes insensitive to frequency drifts of both modes, 
allowing for a narrow band detection of the 
up--converted signal produced by the cavity wall modulation.}
\end{enumerate}
 
\subsection{The rf control loop}
\label{rf}

\begin{figure}[p]
\begin{center} \mbox{\epsfig{file=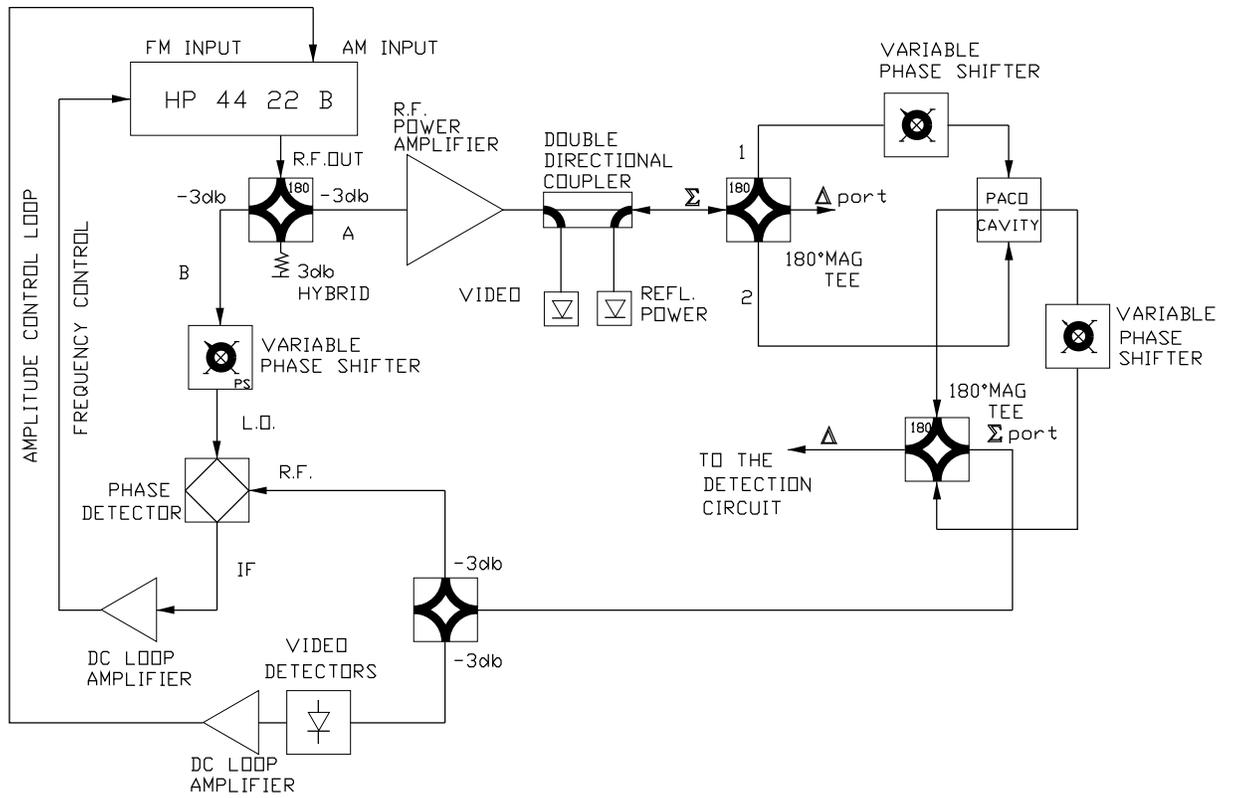}} \end{center}
\caption{rf control loop} 
\label{loop} 
\end{figure}

{  The shematic layout of the rf control loop is shown in fig. \ref{loop}.}

The RF signal generated by the master
oscillator (HP4422B) is fed into the cavity through a TWT 
amplifier giving a saturated output of 20 Watt in the frequency range 2--4 GHz.
The energy stored in the cavity is adjusted at the operating level by controlling the 
output of the master oscillator via the built--in variable attenuator. 

The output signal is divided by a 3 dB power splitter. The $A$ output of the 
splitter is sent to the TWT amplifier, the $B$ output is sent, through the phase shifter (PS), to 
the local oscillator (LO) input of a rf mixer acting as a phase detector (PD).
The output of the rf power amplifier is fed to the resonant cavity through a 
double directional coupler, and a $180^{\rm o}$ hybrid ring acting as a magic--tee.
The rf power enters the magic--tee via the sum arm, $\Sigma$, and is split in two signals 
of same amplitude and zero relative phase, coming out the tee co--linear arms 1 and 2.
 
The rf signal, reflected by the input ports of the cavity, enters the magic--tee 
through the co--linear arms. The two signals are added at the $\Sigma$ arm and sampled by the 
directional coupler to give information about the energy stored in cavity allowing for 
the measurement of the coupling factor, quality factor, stored energy. 
While driving the cavity on the symmetric mode no reflected signal is shown at the $\Delta$ port of 
the magic--tee where the signals coming from the co--linear arms are algebraically added 
to zero due to the $180^{\rm o}$ phase shift.

To get the maximum of the performances of the magic--tee we need to have equal 
reflected signals (phase and amplitude) at the cavity input ports.
To preserve the signal integrity we use matched lines (in phase and 
amplitude) inside the cryostat.
Because the phase shift is very sensitive to temperature inhomogeneities between 
the two cables and the phase difference between the two co--linear arms of the magic--tee 
gives a quite strong signal at the $\Delta$ port, we need to compensate for 
differential thermal contractions of the cables inside the cryostat, leading to phase 
unbalance in the feed lines.
To do that we insert a phase shifter in one of the lines to compensate for differences and to reduce to a minimum the 
leakage of the unwanted modes on the two ports.
As we will show in section \ref{sec:sensit}, mode leakage of the symmetric mode to the 
$\Delta$ port sets a limit to the system sensitivity increasing the overall noise level of the 
detector.

Mode leakage of the antisymmetric mode to the $\Sigma$ port reduces the system sensitivity by 
reducing the signal level available for detection.
The output ports of the cavity are coupled for a maximum output signal 
on the antisymmetric mode (detection mode) and the magic--tee is used to reject the rf power at 
the frequency of the symmetric mode.
A fraction of the signal at the $\Sigma$ port is fed to the rf input of the phase detector 
PD via a low noise rf amplifier.
The intermediate frequency  (IF) output of the phase detector PD is fed back to 
the rf master oscillator to lock the output signal to the resonant frequency of the 
resonator.
The total phase shift around the loop is set through the phase shifter PS, to have the 
maximum energy stored in the detector.
A careful design of the servo loop amplifier (SLA) guarantees the stability of 
the system and the rejection of the residual noise of the master oscillator up to one 
MHz.
The same fraction of the $\Sigma$ output of the output magic--tee is used to keep 
constant, to 100 ppm, the energy stored in the cavity feeding back an error 
signal to drive the electronically controlled output attenuator of the master oscillator.

Great deal of care is needed in tailoring the frequency response of both controls 
because the two loops can interact producing phase--amplitude oscillations in the rf 
fields stored in the cavities.

\subsection{Sensitivity enhancement using the mode symmetry}
\label{sec:sensit}

\begin{figure}[p]
\begin{center} \mbox{\epsfig{file=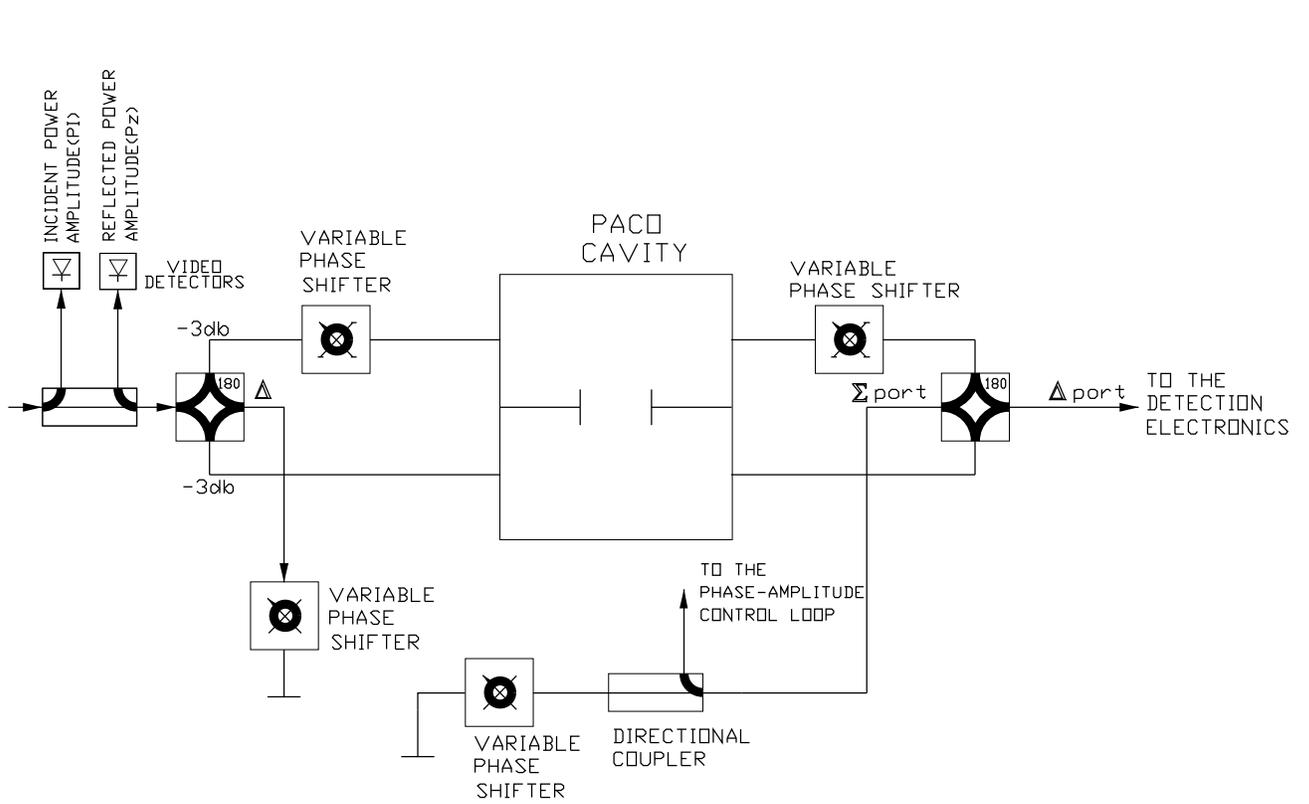}} \end{center}
\caption{Sensitivity enhancement circuit} 
\label{sensit} 
\end{figure}

{  The layout of the sensitivity enhancement circuit is shown in fig. \ref{sensit}.}

The two modes of the detector cavity have (as in the case of two coupled 
pendula) opposite symmetries of the fields. 

Using two separate sets of ports to drive the cavity and to receive the up--converted
signal at the frequency of the antisymmetric mode, the cavity acts as a very sharp 
filter (due to the high $Q$), with an high rejection of the noise coming from the master oscillator at 
the frequency of the up--converted signal. 
This already low residual noise, can be even more reduced in our scheme using two magic--tees to drive
the cavity purely in the symmetric mode and to detect only the up--converted energy, rejecting the unwanted
field components by an amount given  by the magic--tee 
insulation.

In the case of an ideal magic--tee the mode rejection is infinite. 
If the cavity is driven purely in the symmetric mode no symmetric mode 
component is transmitted through the system and there will be no signal at the output port.
In the ideal case this result is obtained also in the more simple scheme used 
by Melissinos and Reece\cite{rrm1,rrm2}, measuring the up--converted power coming out of the detector 
along the input lines.

Our scheme gives better sensitivity and performances in the real case.
The first obvious gain is the sum of the $\Delta$ and $\Sigma$ port insulation of the two tees, plus 
the possibility of adjusting separately the input and output lines to get better mode 
rejection. In a commercial magic--tee the insulation is specified to be 
$\approx 25$ -- $35$ dB over its own bandwidth.
The reason for this quite low insulation is mainly due to the difficulty of balancing 
on a large range of frequency the phases of the signals coming from the two co--linear 
arms of the tee.

A phase unbalance as small as five degrees reduces the insulation from $\Delta$ to $\Sigma$ 
port to only 25 dB.

Our electronic scheme allows for an independent compensation of the magic--tee 
phase mismatch both at the feed frequency and at the detection frequency in a flexible 
way: the phase mismatch is compensated using a variable phase shifter at the input 
of one of the co--linear arms.
The optimum phase at the input side results in a pure excitation of the 
symmetric (drive) mode, keeping the power at the frequency of the 
antisymmetric (detection) mode 70 dB below the level of the drive mode.
Adjusting the phase at the output will couple only the antisymmetric mode 
component, rejecting the symmetric mode component by 70 dB.
The total symmetric--antisymmetric mode rejection of the system is the sum of the attenuation we 
can obtain from the two $180^{\rm o}$ hybrids.

The input and output ports of the two cell cavity need to 
be critically coupled ($\beta$ = 1) to the rf source and to the rf detection system.
In this way we have the optimum transfer of power to the symmetric mode (a 
maximum of stored energy) and to the antisymmetric mode (a maximum in the detector output).
Because the frequency and field distribution of the two modes are quite close, the 
input and output ports are critically coupled to both modes. For that reason 50$\%$
of the symmetric mode signal is coupled to the idle $\Sigma$ port at the output magic--tee, 
and symmetrically 50$\%$
of the antisymmetric mode signal is coupled to the idle $\Delta$ port at the input magic--tee.
{  We remark that, since those two ports are not used in our detection scheme, it is necessary to
reflect back into the cavity the energy flowing out of them. This task is achieved by a proper termination
of the ports; a careful analysis showed that closing the ports with an open circuit completely
decouples the input and output arms (optimum rejection of the symmetric mode)
and maximizes the stored energy and the detector sensitivity.} 

At the $\Sigma$ port of the detection arm we insert a directional coupler to sample 
a tiny amount of the symmetric mode power coming from the cavity.
This signal is fed into the frequency--amplitude servo loop used to lock 
the master oscillator to the cavity frequency and to keep constant 
the energy stored in the cavity.

\subsection{Detection of the converted signal}

\begin{figure}[p]
\begin{center} \mbox{\epsfig{file=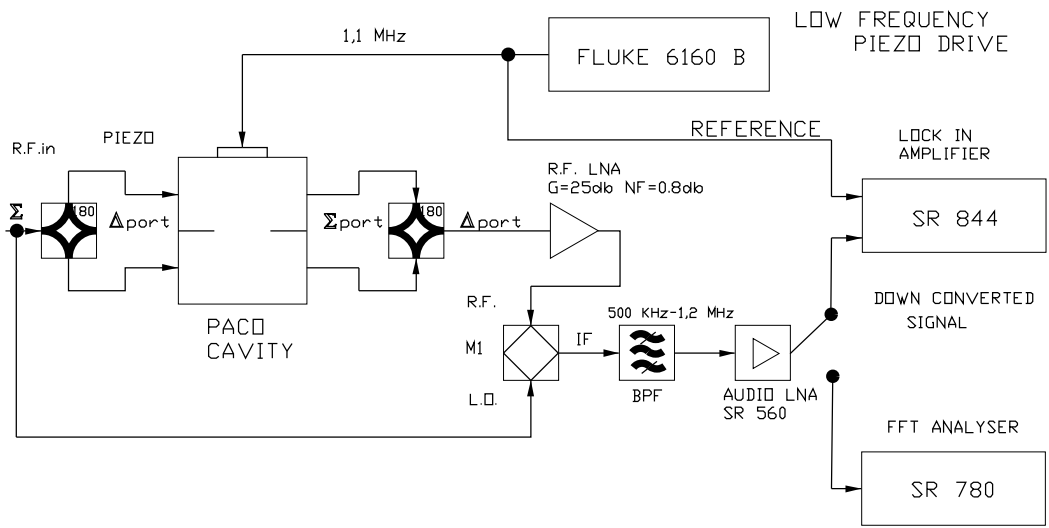}} \end{center}
\caption{Detection system} 
\label{detection} 
\end{figure}

{  The layout of the detection electronics is shown in fig. \ref{detection}.}

The signal converted to the antisymmetric mode by the interaction between the mechanical 
perturbation and the rf fields is coupled to the $\Delta$ port of the detection arm of the rf system 
and amplified by the low noise rf amplifier LNA.
Great deal of care must be used in the choice of the LNA, because the noise figure of 
the amplifier greatly affects the detector's sensitivity in the detection region above the 
mechanical resonances of the cavity (typically spanning the frequency interval 1--10 kHz)
The LNA we choose is a commercial, room temperature, low noise amplifier with 48
dB gain and bandwidth of 500 MHz centered on our operating frequency of 
3 GHz.
The LNA noise figure is 0.6 dB corresponding to a noise temperature $T_n = 340$ K. 

The converted rf signal amplified by the LNA is fed to the rf input port (RF) of a 
low noise double balanced mixer M1; the local oscillator port (LO) of M1 is driven by the 
symmetric mode rf power (at angular frequency $\omega_s$). The LO input level is 
adjusted to minimize the noise contribution from the mixer.
As shown in the previous section the input spectrum to the RF port of the mixer consists of two signals:
the first at angular frequency $\omega_s$ coming from the rf leakage of the symmetric mode through the 
detection system; the second is the converted energy at angular frequency $\omega_a$.
Both signals are down--converted by the M1 mixer giving to the IF port a DC signal 
proportional to the symmetric mode leakage and a signal at angular frequency $\Omega$ proportional to the antisymmetric 
mode excitation.

The down--converted IF output is further amplified using a low noise audio 
preamplifier (Stanford Research SR560).
The combination of tunable built in filters of the SR560 amplifier allows us to further 
reject (if necessary) the DC component coming from the symmetric mode leakage.
Finally the output of the audio amplifier is fed to the lock--in amplifier (Stanford 
Research SR844) used as synchronous detector driven by the low frequency synthesizer used 
to drive the detector cavity through a piezoelectric ceramic at angular frequency $\Omega$.

For the detection electronics of mechanically coupled interactions, at angular frequency $\Omega$, as 
gravitational waves, since the exact frequency and phase of the driving source is not known, we 
can't perform a synchronous detection. We need to perform an auto--correlation of the detector output, 
or to cross--correlate the outputs of two different detectors.
The output of the SR560 preamplifier is fed to an fft signal 
analyzer (Stanford Research SR780) able to average the input signal on a bandwidth as small 
as 0.1 mHz.

The outlined detection scheme gives the benefit of being insensitive to perturbations 
affecting in the same way the frequency of the two modes.

\section{Experimental Results}
\label{sec:results}

The electromagnetic properties were measured in a vertical cryostat after 
careful tuning of the two cells frequencies. In fact in order to get maximum sensitivity we need to have two {\it identical} coupled resonators, or, in other words, a flat field distribution between the two cavities.

The symmetric mode frequency was measured at 3.03 GHz and the mode 
separation was 1.38 MHz.

\begin{figure}[p]
\begin{center} \mbox{\epsfig{file=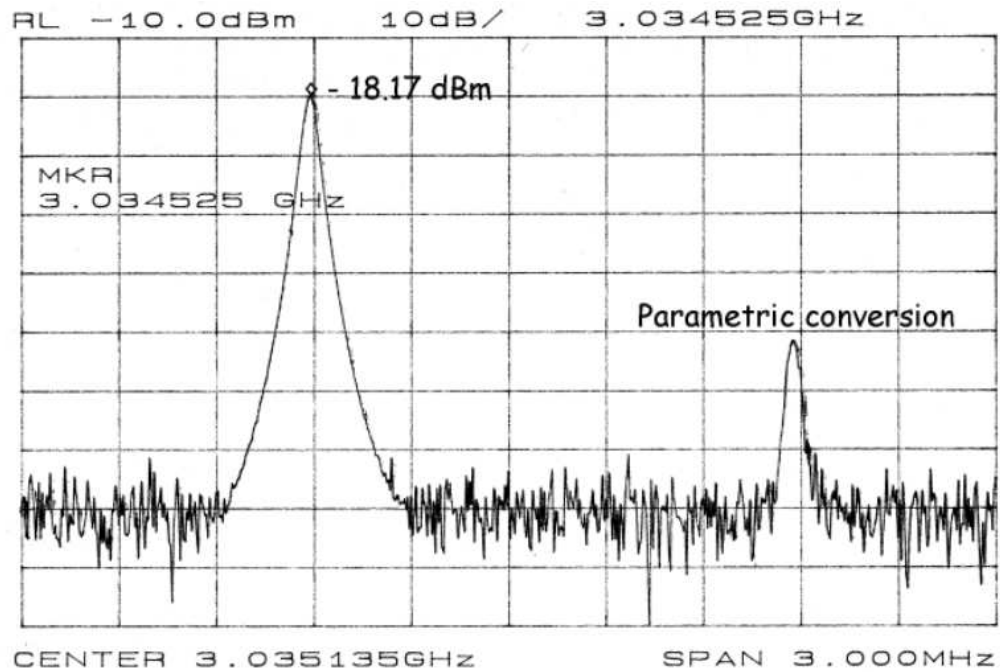,height=9cm}} \end{center}
\caption{Transmission of the symmetric mode (no optimization) {  measured at the $\Delta$ port of the output magic--tee}} 
\label{mis1} 
\end{figure}

\begin{figure}[p]
\begin{center} \mbox{\epsfig{file=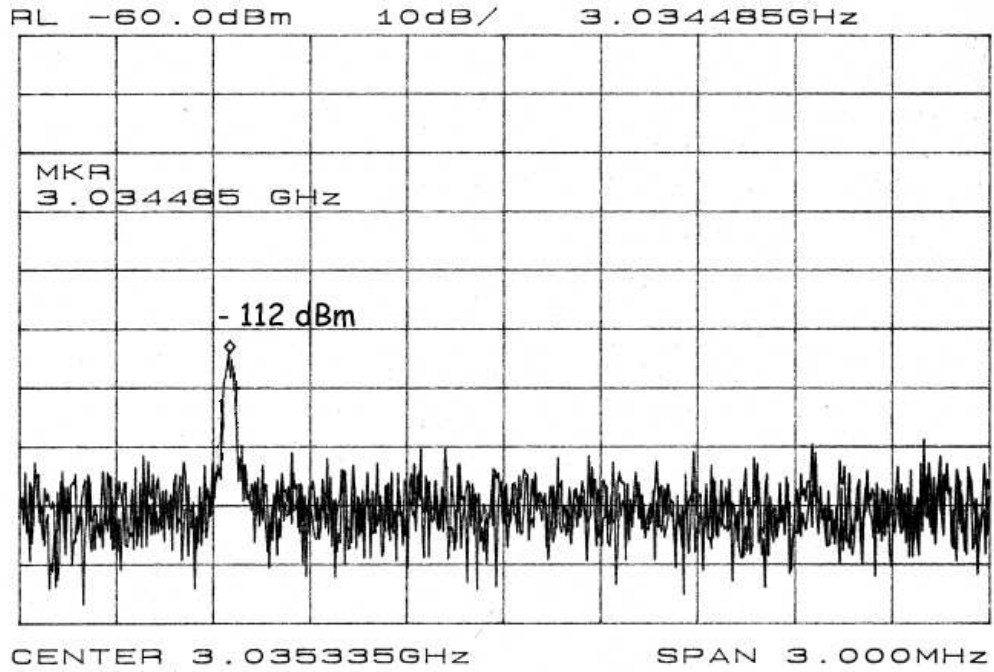,height=9cm}} \end{center}
\caption{Transmission of the symmetric mode in the optimized system,
{  measured at the $\Delta$ port of the output magic--tee. Measurement taken with 1 kHz resolution bandwidth.}} 
\label{mis3} 
\end{figure}

In figure \ref{mis1} the signal from the $\Delta$ port of the output magic--tee is shown for an input power $P_i = 1$ W and no adjustments made on the phase and amplitudes of the rf signal entering and leaving the 
cavity. The overall attenuation of the symmetric mode is $R \approx -48$ dB. 
After balancing the arms of the two magic--tees in order to launch the 
symmetrical mode at the cavity input and to pick up the antisymmetrical one 
at the cavity output, with 1 W (30 dBm) of power at the $\Sigma $ port of the 
first magic--tee, $6.3 \times 10^{-15}$ W (-112 dBm) were detected at the $\Delta $ port of the second 
one, giving an overall attenuation of the symmetric mode of $R \approx -140$ dB (see figure \ref{mis3}).
At a detection frequency of $\Omega / 2 \pi \approx 1$ MHz system sensitivity is quite independent 
from the value of $R$, because of the high cavity $Q_L$. 
Nevertheless for lower frequencies, in a range $\Omega \leq 10$ kHz, where astrophysical sources of 
gravitational waves are expected to exist, this noise source can become dominant and the achieved 
rejection is fundamental in order to pursue the design of a working g.w. detector in the 1--10 kHz frequency range.    

The cavity loaded quality factor was $Q_L = 1 \times 10^9$ at 1.8 K,
and the energy stored in the cavity with 10 W input power was approximately 1.8 J 
(limited by the maximum power delivered by the rf amplifier), 
with both the input and output ports critically coupled ($\beta_1 \approx \beta_2 \approx 1$).

To excite the antisymmetric mode a piezoelectric crystal (Physik Instrument PIC 140, with longitudinal piezoelectric coefficient $\kappa_\ell = 2 \times 10^{-10}$ m/V) has been fixed to one cavity wall. The driving signal to the crystal was provided by a synthesized oscillator with a power output in the range 2--20 mW (3--13 dBm). The oscillator output was further attenuated to reduce the voltage applied to the piezo by a series of fixed attenuators and a variable attenuator (10 dB step). The oscillator frequency was carefully tuned to maximize the energy transfer between the cavity modes.
The signal emerging from the $\Delta $ port of the output magic--tee was 
amplified by the LNA (48 dB gain) and fed into a spectrum analyzer. In figure \ref{mis1} an example of the parametric conversion process is shown. 

The minimum detected noise signal level at the antisymmetric mode frequency, {  with no excitation coming from the piezo}, was $P_{out}(\omega_a) = 5 \times 10^{-19}$ W in a bandwidth $\delta f = 100$ Hz, giving a noise power spectral density ${\mathcal P}_{out}(\omega_a) = 5 \times 10^{-21}$ W/Hz; the main contribution to this signal was the johnson noise of the rf amplifier used to amplify the signal picked from the $\Delta$ port of the output magic--tee.

Taking into account the input and output coupling coefficients system sensitivity is given by\cite{rrm1}:
\be
h_{min} = \frac{\delta \ell}{\ell} = \frac{1}{Q_L}\left[ \frac{{\mathcal P}_{out}(\omega_a)}{P_{in}(\omega_s)}
\frac{(1+\beta_1)(1+\beta_2)}{\beta_1\beta_2} \right]^{1/2} \approx 3 \times 10^{ -20} \mathrm{(Hz)}^{-1/2} 
\ee

The experimental results are summarized in table II.

\section{Future Plans}
\label{future}

In order to design a working g.w. detector based on the parametric converter principle, we must concentrate on the following main items:
\begin{enumerate}
\item {{\em Lower detection frequency}. The test made at 1 MHz have to be repeated at lower frequencies 
(say in the range between 1--10 kHz) to span the frequency region of interest for astrophysical sources.} 
\item{{\em Tuning system}. Since our detector might be particularly useful for the analysis of the gravitational signal coming from a continous source of known frequency, a tuning system has to be designed for the fine frequency adjustment needed to match the detection frequency with the source. One possible solution is to mount a superconducting bellow in the region of the coupling between the two cavities to vary the distance (i.e. the coupling strength) between them\cite{nota4}. Even if this solution appears to be feasible, at least in principle, some tests are needed to verify its applicability; in particular the tuning system should not degrade too much the cavity $Q$. Extensive numerical simulations are at the moment underway.}
\item {{\em System stability} over time ranges of the order of magnitude of 
the inverse of our detection bandwidth. Obviously if we are going to detect 
our signal in a narrow bandwidth, we have to check carefully that system 
parameters do not change significantly over a time scale as long as 
possible. In particular the most critical parameter is the frequency stability of the system.}
\item {Among other possible improvements, a {\em cryogenic preamplifier}, with operating temperature
in the 20--30 K range, will be implemented.}
\end{enumerate}

At the end of the experimental tests on the cylindrical detector, we shall 
build and test the spherical cavities detector configuration, which should 
give best performances in view of gravitational waves detection, due to the 
more favourable electromagnetic geometric factor and field symmetries. Moreover, using niobium 
sputtered copper cavities, we should take advantage of the very high quality 
factors obtained so far with this technique (up to $10^{11}$ in 1.5 GHz structures)\cite{benve}.

If the goal sensitivity would be obtained, a detector based on these principles and with a suitable geometry could be an 
interesting candidate in the search of gravitational waves. In fact it is conceivable to push down the 
rf operating frequency around 300 MHz and the detection frequency in the 1--10 
kHz range. Making a series of similar detectors, working at different 
frequencies and/or mode separations we can cover the 10$^{3}$--10$^{6}$ Hz 
range.

Recent works\cite{theory1,giovannini} focused their attention on 
the dection of stochastic g.w., in particular of the relic g.w. 
background, and pointed out that a relic background detected at high 
frequency would be unambiguously of cosmological origin. The detection of 
stochastic g.w. could be done by correlating two (or more) detectors 
put at a distance small compared to the g.w. wavelength (so that the signals 
could be correlated) but large enough to be sufficient to decorrelate local 
noises. With this experimental arrangement system sensitivity could be 
increased by several orders of magnitude\cite{maggiore} making possible the detection of very low 
signal levels.



\section{Acknowledgements}

We wish to thank dr. S. Farinon of INFN--Genova for the contribution given in the development and analysis of the detector mechanical model, dr. C. Benvenuti and dr. E. Chiaveri, with the staff of CERN, for the chemical and thermal treatments of the niobium cavity. We had numerous fruitful discussions with many of our colleagues, in particular with dr. A. Chincarini.

        \bibliographystyle{prsty}
        \bibliography{038105RSI}

\begin{thebibliography}{10}
\bibitem{nota1}
In this section were not differently specified we take $\omega_s \approx \omega
  _{a} = \omega $.
\bibitem{nota4}
We remind that the detection frequency, i.e. the frequency split between the
  two modes of the coupled system, is proportional to the coupling coefficient.
\end{thebibliography}

\newpage

\begin{table}[p]
\label{tab:modes}
\caption{Mechanical resonances spectrum}
\begin{tabular}{cc}
Mode Number & Frequency (Hz) \\
\tableline
1 & 633.15 \\
2 & 654.48 \\
3 & 713.96 \\
4\tablenote{Modes 4--11 are doubly degenerate} & 1313.4 \\
5 & 1318.5 \\
6 & 1412.2 \\
7 & 1606.6 \\
8 & 1808.4 \\
9 & 1826.0 \\
10 & 1950.9 \\
11 & 2018.2 \\
\end{tabular}
\end{table}

\begin{table}[p]
\label{tab:meas}
\caption{Experimental results}
\begin{tabular}{lcc}
Symmetric mode frequency & $\omega_s/2\pi$	&	3.0343 $\times 10^9$ Hz 	\\
Antisymmetric mode frequency & $\omega_a/2\pi$	&	3.0357 $\times 10^9$ Hz		\\
Mode splitting & $\Omega/2\pi$		&	1.38 $\times 10^6$ Hz		\\
Input power & $P_{in}(\omega_s)$&	18 W \\
Input coupling coefficient & $\beta_1$ & 0.92 \\
Output coupling coefficient & $\beta_2$ & 1.00 \\
Loaded quality factor of the symmetric mode & $Q_{sL}$ & 1.01 $\times 10^9$ \\
Loaded quality factor of the antisymmetric mode & $Q_{aL}$ & 9.92 $\times 10^8$ \\
Stored energy & $U_s$ & 1.8 J \\
Noise power spectral density & ${\mathcal P}_{out}(\omega_a)$ & 5 $\times 10^{-21}$ W/Hz \\
Minimum measurable relative displacement & ${h}_{min}$ & 3.3 $\times 10^{-20}$ Hz$^{-1/2}$ \\
\end{tabular}
\end{table}



\newpage

\newpage

\newpage

\newpage

\newpage

\newpage

\newpage

\newpage

\newpage

\newpage

\newpage

\end{document}